\documentclass[aps,pra,twocolumn,groupedaddress]{revtex4}
\usepackage{amsmath}
\usepackage{graphicx}
\usepackage{amsfonts}
\usepackage{color}
\makeatletter
\def\captionof#1#2{{\def\@captype{#1}#2}}
\makeatother
\begin{document}

\title{Non-linear transport in an out-of-equilibrium single-site Bose Hubbard model: scaling, rectification and time dynamics} %

\author{Archak Purkayastha}
\affiliation{International centre for theoretical sciences, Tata Institute of Fundamental Research, Bangalore - 560089, India}
\author{Abhishek Dhar}
\affiliation{International centre for theoretical sciences, Tata Institute of Fundamental Research, Bangalore - 560089, India}
\author{Manas Kulkarni}
\affiliation{International centre for theoretical sciences, Tata Institute of Fundamental Research, Bangalore - 560089, India}

\begin{abstract}

Recent experiments in hybrid-quantum systems facilitate the potential realization of one of the most fundamental interacting Hamiltonian-Reservoir system, namely, the single-site Bose-Hubbard model coupled to two reservoirs at different temperatures. Using Redfield equations in Born-Markov approximation, we compute non-equilibrium average particle number, energy and currents beyond linear response regime, both time-dynamics and steady state and investigate its dependence on various tunable parameters analytically. We find interesting scaling laws in high temperature regimes that are independent of choice of bath spectral functions. We also demonstrate that the system shows very interesting particle and energy current rectification properties which can be controlled via the relative strength of interaction  and temperatures, as well as via the degree of asymmetry in system-bath coupling. Specifically, we find inversion of direction of energy rectification as a function of the relative strength of the interaction strength and the temperatures. We also show that, in the limit of low-temperature and high interaction strength our results are consistent with the non-equilibrium spin-Boson model. Our results are experimentally relevant not only to hybrid quantum systems, but also in other areas such as molecular junctions.  
\end{abstract}

\maketitle

\subsection{\label{sec:A} Introduction}

Far-from-equilibrium systems beyond the paradigm of linear response has been a subject of growing theoretical and experimental interest \cite{sskj,qh1,qh2,qh3,qh4,qh5,KLH2015,ah0}. Such studies are important from fundamental perspective and from the point of view of device applications. Especially, understanding the role of interactions is of great interest. Important fundamental problems to explore include, role of far-from-equilibrium physics in physical quantities of interest and the intricate interplay between system and bath degrees of freedom when interactions are involved. Some of the main consequences of interactions range from fundamental physics (such as the phenomenon of Kondo effect \cite{kontosPRL,keijiPRL}) to device applications (such as diodes/rectifiers\cite{kul2015,spins_diode1,spins_diode2}). Even though the role of interaction in device applications in fermionic systems is well studied \cite{Dubi2015,Chang2010,Ono2002}, the corresponding bosonic set-ups are much less explored. However, with the advent of the relatively new field of photonics, bosonic set-ups for device applications such as a potential optical diode are gaining popularity \cite{Wang2013, Shen2014, Mascarenhas2014}.  With recent cutting-edge technology in hybrid quantum systems, it is now possible to design interacting bosonic Hamiltonians and reservoirs \cite{ber1,ah0,ah1,ah2,ah3,ah4,ah5}.  There is progress not only in fabricating Hamiltonian-Reservoir systems, but also, great advance in measuring physical quantities of interest such as photon number, photon statistics \cite{petta0} and photon current \cite{Google2016} .

One of the most basic interacting bosonic Hamiltonian system that one can think of is a single bosonic site with Bose-Hubbard  interaction, hereafter, called the single-site Bose Hubbard (SSBH) model. In optics, such interaction is often called the Kerr interaction strength. In the field of hybrid quantum systems a Hamiltonian with such an interaction can be potentially experimentally realized in more than one way. The Jaynes-Cummings model, an experimentally realized light-matter system, can be tuned to the dispersive regime where it behaves like a SSBH model. The photon-spin interaction in this limit can be integrated out ``pertubatively'' to generate a non-linear Bose-Hubbard like interaction between photons \cite{Blais2009,ah1}. Alternatively there are interesting potential realizations of Bose-Hubbard interactions between photons which involve 4-level atoms in an optical cavity \cite{Hartman2007,Ling2011}. These realizations offer large tunability of parameters. Specifically, while the former realization involving Jaynes-Cummings model has a small interaction strength compared to the linear term, the later realizations have very large interaction strength.

Another area of applicability for such interacting bosonic Hamiltonians, which is perhaps more suited to non-equilibrium measurements, are the fields of molecular thermoelectrics and nanophononics. These fields typically deal with molecular junctions connecting two reservoirs, experimentally, which are often two large chemical compounds \cite{exp1,exp2,exp3,exp4,exp5,exp6}. Role of interactions in phononic transport through such systems is of interest both experimentally and theoretically\cite{ds1}.    

There has been a large amount of work on such a SSBH model with a finite interaction strength coupled to a single bath \cite{Walls1980,Haake1986,Milburn1986,Haake1987,Alicki1989,Kartner1993}. On the other hand, non-equilibrium spin boson (NESB) model, which corresponds to the limit of very large interaction strength, has been well studied in out-of-equilibrium set-ups \cite{SegalPRL2005, SegalJChemPhys2005, SegalPRB2006, NESBcond1, keijiPRL, NESBcond2} and is of growing experimental significance. The conductance of an anharmonic junction with quartic anharmonicity has been studied recently \cite{jt1,jt2}. 
However, there is essentially no investigation of the SSBH model with finite interaction strength in a far-from-equilibrium setup via connection with multiple reservoirs. In this paper, we investigate the SSBH model weakly coupled to two bosonic baths at different temperatures beyond the linear response regime.

The organization of our paper is as follows: In section~\ref{sec:B}, we describe our setup. In section~\ref{sec:C}, we give the Redfield quantum master equation for our setup and in section~\ref{sec:D}, we find the non-equilibrium steady state (NESS) properties.  In section~\ref{sec:E}, we look at scaling behaviour of average particle number and energy at NESS and in section~\ref{sec:F}, we discuss the scaling behaviour and rectification of NESS particle and energy currents.  In section ~\ref{sec:TD}, we present results for time dynamics of various physical quantitites which has become of growing recent interest. Finally, in section~\ref{sec:G}, we summarize our main results along with an outlook. 

\subsection{\label{sec:B} Model and Setup}
We consider a single site with Bose-Hubbard interaction connected to two bosonic baths in non-equilibrium. The baths are taken to be quadratic and the system-bath couplings are taken to be bilinear. Thus our set-up is given by the full system+bath Hamiltonian 
\begin{align}
&\hat{H} = \hat{H}_S+\hat{H}_B+\hat{H}_{SB} \nonumber\\
&\hat{H}_S = \Omega_0 \hat{a}^\dagger\hat{a} + \chi (\hat{a}^\dagger\hat{a})^2 \nonumber\\
&\hat{H}_B = \sum_{\ell=1}^2\sum_{r=1}^\infty \Omega_r^\ell \hat{B}_r^{\ell \dagger} \hat{B}_r^\ell~, \\
&\hat{H}_{SB} = \varepsilon\sum_{\ell=1}^2 \sum_{r} (\kappa_{\ell r} \hat{B}_r^{\ell \dagger} \hat{a} + \kappa_{\ell r}^* \hat{a}^{\dagger} \hat{B}_r^\ell)~,\nonumber   
\end{align}
where $\hat{a}$ correspond to bosonic annihilation operators and  $\hat{B}_{r}^{\ell}$ to those of $\ell$th bath ($\ell=\{1,2\}$).  Note that  $\hat{N}=\hat{a}^\dagger\hat{a}$ is a conserved quantity with respect to the system Hamiltonian $H_S$. The energy spectrum of the system Hamiltonian can be easily written down and it has the non-linear form $\hat{E}=\Omega_0 \hat{N} + \chi \hat{N^2}$. 

The baths are quadratic and have infinite degrees of freedom. $\varepsilon$ is a dimensionless parameter that controls system bath coupling. We assume that, initially, there is no coupling between the system and the baths, and the two baths are at thermal equilibrium with their own inverse temperature $\beta_1$ and $\beta_2$ and chemical potential $\mu_1$ and $\mu_2$. Thus the  initial bath correlation functions satisfy the thermal properties :
\begin{equation}
\label{initial_bath_corr}
\begin{split}
\langle\hat{B}_r^\ell\rangle = 0 , \hspace{2pt} <\hat{B}_r^{\ell\dagger} \hat{B}_s^\ell >_B = \mathfrak{n}_{\ell}(\Omega_r^\ell)\delta_{r s }~,  
\end{split}
\end{equation}
where $\mathfrak{n}_\ell(\omega)= [{e^{\beta_{\ell}( \omega-\mu_{\ell})}- 1}]^{-1}$ is the bosonic distribution function. We also introduce the bath spectral functions:
\begin{equation}
\mathcal{J}_{\ell}(\omega)=2\pi \sum_r  \mid \kappa_{\ell r} \mid^2 \delta(\omega - \Omega_r^\ell)~. 
\end{equation} 

We will assume, 
\begin{align}
\mathcal{J}_\ell(\omega) = \Gamma_\ell \mathcal{J}(\omega)~, 
\end{align}
i.e, the two baths have same density of states, but the system-bath coupling is different in general. $\Gamma_\ell$ has dimensions such that $\mathcal{J}_\ell(\omega)$ has dimensions of energy.  
   
The quantitative nature of the results will depend on our choice of spectral function. We assume a general spectral function which is commonly used in bosonic systems:
\begin{align}
\label{J_general}
\mathcal{J}(\omega)=\omega^s e^{-\omega/\omega_c}~,
\end{align}
where $\omega_c$ gives the cut-off frequency. The cut-off frequency is considered very large so that the system energy levels near the edge of the bath spectrum correspond to extremely high energies, which do not really contribute to the system properties at the chosen set of temperatures and chemical potentials. This is satisfied when $\omega_c \gg \Omega_0, \chi, \beta_1^{-1}, \beta_2^{-1},\mu_1,\mu_2$. Also, we are concerned with a photonic or phononic system, so we will set $\mu_1=\mu_2=0$ finally.

\subsection{\label{sec:C} The Quantum Master Equation (QME)}

We want to investigate this set-up without any restrictions on the interaction strength $\chi$. So, we adopt the method of QME under Born-Markov approximation, which is only valid when system-bath coupling is weak, and gives results only to leading order in system-bath coupling. The microscopically derived Redfield QME for our system coupled to single bath was first written down in \cite{Haake1986}. It is straighforward to generalize to two baths. Let us define $\hat{\rho} \equiv Tr_B(\hat{\rho}_{full})$ with  $\hat{\rho}_{full}$ being the full density matrix of system+bath and $Tr_B(..)$ implying trace taken over bath degrees of freedom. The  weak-coupling Redfield equation for the density matrix of the system $\hat{\rho}$ of our set-up is given by :

\begin{align}
\label{rqme}
&\frac{\partial \hat{\rho}}{\partial t} = i[\hat{\rho},\hat{H}_S] - \varepsilon^2 \Big ( [\hat{\rho} F(\hat{\tilde{\omega}})\hat{a},\hat{a}^\dagger]  + [\hat{a}^\dagger, G(\hat{\tilde{\omega}})\hat{a} \hat{\rho}] + h.c. \Big)~,
\end{align}
where
\begin{align}
\label{FG}
&\hat{\tilde{\omega}} = \Omega_0+\chi(2\hat{N}+1), \nonumber \\
&F(\hat{\tilde{\omega}}) = F_1(\hat{\tilde{\omega}})+F_2(\hat{\tilde{\omega}}), \hspace{5pt} G(\hat{\tilde{\omega}}) = G_1(\hat{\tilde{\omega}})+G_2(\hat{\tilde{\omega}})  \\
& F_\ell(\hat{\tilde{\omega}}) = \frac{1}{2} \mathcal{J}_\ell(\hat{\tilde{\omega}})\mathfrak{n}_\ell(\hat{\tilde{\omega}}) - i \mathcal{P}\int_{-\infty}^{\infty}\frac{d\omega}{2\pi} \frac{\mathcal{J}_\ell(\omega)\mathfrak{n}_\ell(\omega)}{\omega-\hat{\tilde{\omega}}} \nonumber\\
& G_\ell(\hat{\tilde{\omega}}) = \frac{1}{2} \mathcal{J}_\ell(\hat{\tilde{\omega}})(\mathfrak{n}_\ell(\hat{\tilde{\omega}})+1) - i \mathcal{P}\int_{-\infty}^{\infty} \frac{d\omega}{2\pi}\frac{\mathcal{J}_\ell(\omega)(\mathfrak{n}_\ell(\omega)+1)}{\omega-\hat{\tilde{\omega}}} \nonumber
\end{align}
and $h.c.$ stands for Hermitian conjugate. Note that the Redfield equation assumes $\varepsilon \ll 1$ and keeps terms only upto $O(\varepsilon^2)$, i.e, only upto quadratic in system-bath coupling (the Born approximation). In deriving the above equation, we have also done the Markov approximation which assumes that the observation time is much larger than the time scale of relaxation of bath correlation functions.   To ensure weak system-bath coupling, we choose $\varepsilon=0.1$ and $\kappa_{\ell r}$ to be of the same order as $\Omega_0$.

The Markov assumption entails that Eq.~\ref{rqme} is not valid at short times. In fact, as pointed out in Ref \cite{Alicki1989}, this equation is not completely positive and may lead to unphysical states at small times for certain initial conditions. A way around suggested in Ref \cite{Alicki1989} is to derive Lindblad equation in terms of eigenbasis operators of $\hat{H}_S$ via the secular approximation. The equation so derived respects complete positivity at all times. In the following, we will be interested only in the diagonal elements of $\hat{\rho}$ in the eigenbasis of $\hat{H}_S$. It turns out that the evolution equation for these elements as derived from the Redfield equation and from the eigenbasis Lindblad equation are exactly the same.

In equilibrium, i.e, when $\beta_1=\beta_2=\beta$, $\mu_1=\mu_2=\mu$, it can be checked by direct substitution that the thermal state
\begin{align}
\label{rho_eq}
\hat{\rho}_{eq} = \frac{e^{-\beta (\hat{H}_S - \mu \hat{N}})}{Z}
\end{align}
is the steady state of the  Eq.~(\ref{rqme}), where $Z~=~Tr\big [e^{-\beta (\hat{H}_S - \mu \hat{N}}\big]$ is the equilibrium partition function.

\subsection{\label{sec:D} The Non-equilibrium Steady State (NESS)}

To calculate various physical observables in NESS beyond linear response,  we need to find the NESS density matrix. Also, in non-equilibrium, since we have no guess for direct substitution, we need to find the $\rho$ directly from Eq \ref{rqme}. Since both the number operator $\hat{N}$ and the system Hamiltonian $H_S$ are diagonal in occupation number basis, the NESS transport properties as well as average occupation and energy of the system  can be found from the steady state diagonal elements of $\hat{\rho}$ in this basis. The occupation number basis satisfies 
\begin{align}
\hat{a}|n\rangle = \sqrt{n} |n-1\rangle, \hspace{5pt} \hat{a}^\dagger|n\rangle = \sqrt{n+1} |n+1\rangle~.
\end{align} 
The  evolution equation for the diagonal elements $\rho_n = \langle n | \hat{\rho} | n \rangle$ is given by 
\begin{align}
\label{rho_n_equation}
\frac{d \rho_n}{dt} = -\varepsilon^2 \left[ \rho_n ( C_n + D_n)- \rho_{n-1}C_{n-1} - \rho_{n+1}D_{n+1} \right]~, 
\end{align}
where
\begin{align}
\label{CD}
&C_n = C^{(1)}_{n}+C^{(2)}_{n}, \hspace{7pt} D_n = D^{(1)}_{n}+D^{(2)}_{n} \nonumber \\
&C^{(\ell)}_{n} = (n+1)\mathcal{J}_\ell(\omega_n)\mathfrak{n}_\ell(\omega_n), \nonumber \\
&D^{(\ell)}_{n} = n\mathcal{J}_\ell(\omega_{n-1})(\mathfrak{n}_\ell(\omega_{n-1})+1) \nonumber \\
&\omega_n = \Omega_0+\chi(2n+1)~.
\end{align}
In the steady state we set the LHS of Eq.~\ref{rho_n_equation}) to zero. This leads to a difference equation. Noting  that $C^{(\ell)}_{-1}=C_{-1} = 0, D^{(\ell)}_{0}=D_0 = 0$, 
 we obtain,  by recursion, the solution
\begin{align}
&\rho_n D_n = \rho_{n-1}C_{n-1} \label{rho_n_condition}\\
&\Rightarrow\rho_n = \rho_0\prod_{p=1}^{n}\frac{C_{p-1}}{D_p} =\rho_0\prod_{p=1}^{n} \frac{\sum_{\ell=1}^2\Gamma_\ell \mathfrak{n}_\ell(\omega_{p-1})}{\sum_{\ell=1}^2\Gamma_\ell ( \mathfrak{n}_\ell(\omega_{p-1})+1)} \nonumber \\
&\textrm{for} \hspace{5pt} n={1,2,3,...}  \label{rho_n}~.
\end{align}
The constant $\rho_0$ is fixed from the normalization condition $\sum_{n} \rho_n=~1$, i.e, trace of density matrix is unity. Thus,
\begin{align}
\label{Z_def}
\rho_0 = \tilde{Z}^{-1} = \Big[1+\sum_{n=1}^{\infty}  \prod_{p=1}^{n} \frac{\sum_{\ell=1}^2\Gamma_\ell \mathfrak{n}_\ell(\omega_{p-1})}{\sum_{\ell=1}^2\Gamma_\ell ( \mathfrak{n}_\ell(\omega_{p-1})+1)}   \Big]^{-1}~,
\end{align}
where $\tilde{Z}$ is a normalization constant and is analogous to the partition function in equilibrium systems. It can be easily checked by putting $\mathfrak{n}_1(\omega)=\mathfrak{n}_2(\omega)=\mathfrak{n}(\omega)$ in Eq.~(\ref{rho_n}) that $\rho_n$ indeed gives Eq.~(\ref{rho_eq}) in equilibrium. Also, in the equilibrium case, Eq.~(\ref{rho_n_condition}) corresponds to the detailed balance condition. In the non-equilibrium case, while this still looks like the detailed balance condition, we note that in general it is not possible to define an effective temperature.  
Also note that $\rho_n$ is independent of the choice of spectral function of the bath. 

\begin{figure}[t]
\includegraphics[scale=0.45]{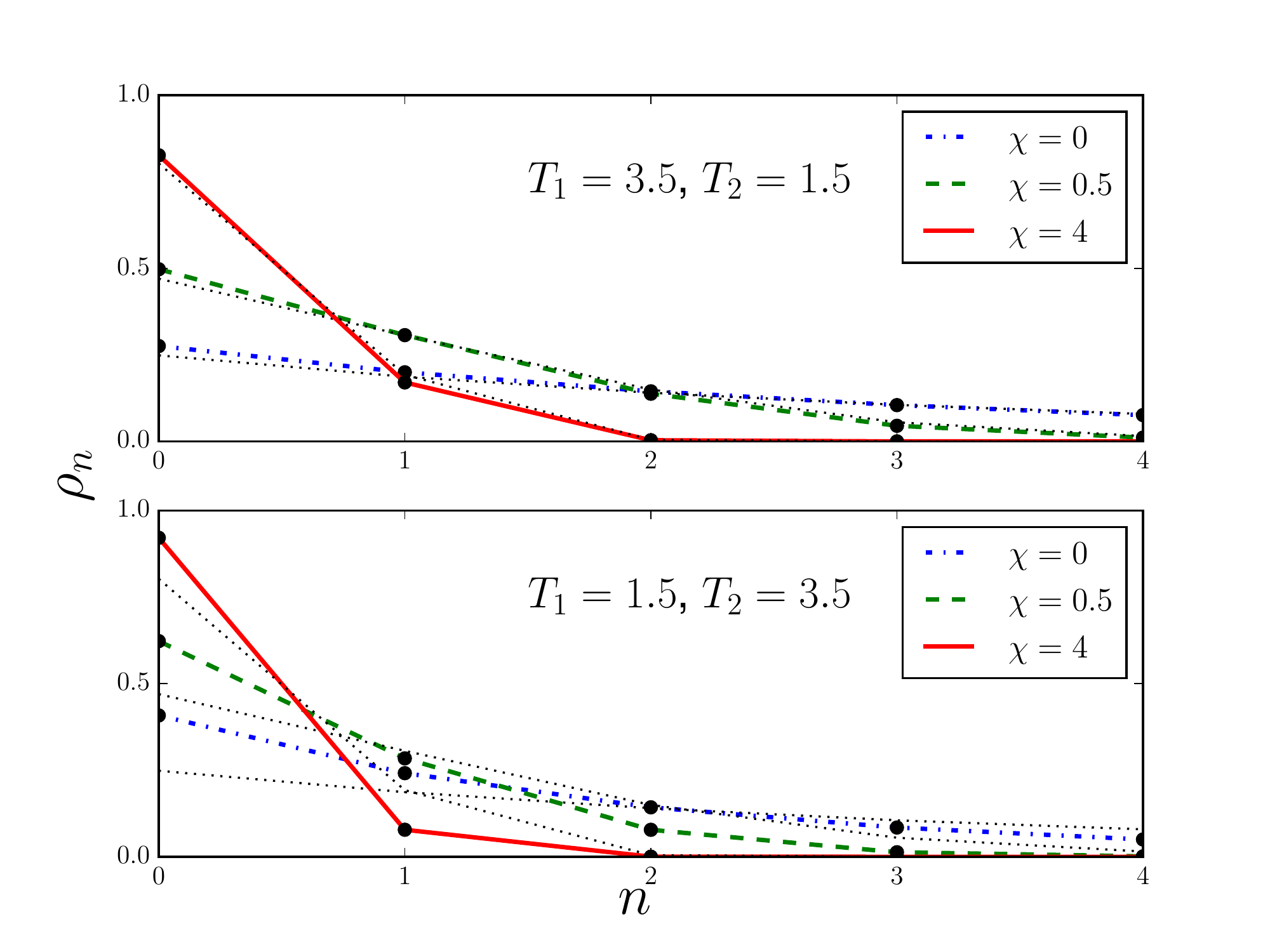} 
\caption{(color online) The plot of population density (diagonal elements of density matrix in eigenbasis of system Hamiltonian) of SSBH model in equilibrium and under thermal bias with assymetric system-bath coupling ($\Gamma_1=0.4,\Gamma_2=1.6$) and $\Omega_0=1$. The top and bottom panels are for interchanged hot and cold baths. The three dotted lines (in each plot) are the corresponding equilibrium distributions ($T_1=T_2=3.5$), i.e, Eq. \ref{rho_eq} for the values of $\chi$ mentioned in the legend. The deviation from equilibrium is more prominent in the bottom panel. For large interaction strength ($\chi=4$), only two levels have non-negligible probability, like a spin-boson model. All energy variables are measured in units of $\Omega_0$.} \label{fig:one_rho_n}
\end{figure} 

The explicit expression for population $\rho_n$ of bosons in NESS, given by Eq.~\ref{rho_n}, is the central result that allows us to go beyond linear response in this interacting bosonic problem. The population, in itself, is a physically measurable quantity and, as we show below, can be used to compute various other physical observables. 

In Fig.~\ref{fig:one_rho_n}, we show the plots of population density $\rho_n$ of the system under asymmetric system-bath coupling. Since system-bath coupling is asymmetric, the out-of-equilibrium distribution changes under interchange of hot and cold baths. For high interaction strength, i.e, for $\chi \gg \Omega_0, T_1,T_2$, only the lowest two levels have non-negligible probability. In this regime, we can truncate the energy spectrum in just two levels. Then $\rho_0$ and $\rho_1$ become  
\begin{align}
\label{rho_SB}
\rho_0 \approx \frac{\sum_{\ell=1}^2 \Gamma_\ell \big(\mathfrak{n}_\ell(\omega_0) +1 \big) }{\sum_{\ell=1}^2 \Gamma_\ell (1+2\mathfrak{n}_\ell(\omega_0))},& \hspace{2pt} 
\rho_1 \approx \frac{\sum_{\ell=1}^2 \Gamma_\ell \mathfrak{n}_\ell(\omega_0) }{\sum_{\ell=1}^2 \Gamma_\ell (1+2\mathfrak{n}_\ell(\omega_0))} \nonumber \\
& \hspace{10pt} \forall \hspace{5pt} \chi \gg \Omega_0, T_1,T_2
\end{align}
with $\omega_0 = \Omega_0+\chi$. The above results are exactly the same as obtained for a non-equilibrium spin-boson model (NESB) by using the Redfield equation \cite{SegalPRL2005}.
Thus for $\chi \gg \Omega_0, T_1, T_2$, the system becomes identical to the NESB. 
Since NESB is already a well explored problem, in the following, we will be mainly interested in the physics beyond this regime. 
We also note that for $\chi=0$ the system reduces to a harmonic oscillator and in this case the population $\rho_n^{\chi=0}$ is given in terms of an effective temperature $T_{eff}=1/\beta_{eff}$, i.e  $\rho_n^{\chi=0} \propto e^{-\beta_{eff} \Omega_0 n}$ with
\begin{align}
\coth (\beta_{eff}\Omega_0)= \frac{\Gamma_1 \coth (\beta_{1}\Omega_0)+\Gamma_2 \coth (\beta_2\Omega_0)}{\Gamma_1+\Gamma_2}~ \label{beta_eff}
\end{align}
which is consistent with the  finding in Ref. \cite{dhar2012}.

\subsection{\label{sec:E} Average occupation and energy}

First we will look at the average occupation and energy of the system. These quantities are measurable in current state-of-the-art experiments in quantum light-matter hybrid systems \cite{ah1,petta0}. The expressions for these are given by
\begin{align}
\label{occ_and_E}
&\langle \hat{N} \rangle  = \sum_{n=1}^{\infty} n \rho_n = \frac{1}{\tilde{Z}}\sum_{n=1}^{\infty} n \prod_{p=1}^{n} \frac{\sum_{\ell=1}^2\Gamma_\ell \mathfrak{n}_\ell(\omega_{p-1})}{\sum_{\ell=1}^2\Gamma_\ell ( \mathfrak{n}_\ell(\omega_{p-1})+1)}~, \nonumber\\ 
&\langle \hat{H}_S \rangle  = \sum_{n=1}^{\infty} E_n \rho_n = \frac{1}{\tilde{Z}}\sum_{n=1}^{\infty} E_n \prod_{p=1}^{n} \frac{\sum_{\ell=1}^2\Gamma_\ell \mathfrak{n}_\ell(\omega_{p-1})}{\sum_{\ell=1}^2\Gamma_\ell ( \mathfrak{n}_\ell(\omega_{p-1})+1)}~,
\end{align}
with $E_n = \Omega_0 n+\chi n^2$.

In NESB limit, these average quantities can be trivially found from Eq.~\ref{rho_SB}. They become $\langle \hat{N} \rangle \approx \rho_1$, and $\langle \hat{H}_S \rangle \approx \omega_0 \rho_1$. We are interested in going beyond the NESB regime. So let us look at the regime of high temperatures $T_1, T_2 \gg \chi,\Omega_0$. First, we look at the normalization constant defined in Eq.~\ref{Z_def}. which can be written as
\begin{align}
&\tilde{Z} = 1+ \sum_{n=1}^\infty \exp\left[-\sum_{p=1}^{n} \log\left(f(\omega_{p-1}) \right)\right] \nonumber \\
&f(\omega_p)=  \frac{\sum_{\ell=1}^2\Gamma_\ell ( \mathfrak{n}_\ell(\omega_{p})+1)}{\sum_{\ell=1}^2\Gamma_\ell \mathfrak{n}_\ell(\omega_{p})} ~.
\end{align}
Note that  $f(\omega_p)>1$, and hence $log\big(f(\omega_p)\big)>0$. It follows that there is a energy level cut-off $n^*$ beyond which the energy levels have negligible contribution to $\tilde{Z}$. For high enough temperatures, we can assume, $\beta_\ell \omega_{p} \ll 1$ $\forall p<n^*$. Under this condition, we can expand $f(\omega_p)$ to obtain (after some amount of algebra),
\begin{align}
 f(\omega_p) &\approx 1+\frac{\omega_p}{\tilde{T}}, \nonumber \\
{\rm hence}~~log\big( f(\omega_p) \big) &\approx log \big(1+\frac{\omega_p}{\tilde{T}}\big) \approx \frac{\omega_p}{\tilde{T}}~,
\end{align}
with 
\begin{align}
\label{Teff}
\tilde{T} = \frac{\Gamma_1 T_1 + \Gamma_2 T_2}{\Gamma_1 + \Gamma_2}~.
\end{align}
The contribution of terms $n> n^*$ is small and so their precise form is irrelevant. Hence we get 
\begin{align}
\tilde{Z}&\approx 1+\sum_{n=1}^{\infty} \exp\left[-\frac{1}{\tilde{T}}\sum_{p=1}^{n} \big( \Omega_0 + (2p-1)\chi \big) \right] 
\nonumber \\
&=\sum_{n=0}^{\infty} \exp\left[-\frac{1}{\tilde{T}}( \Omega_0 n +\chi n^2 )\right]~.
\end{align}
Thus, for $T_1,T_2\gg \Omega_0,\chi$ the normalization constant has the same form as the equilibrium partition function with the effective temperature $\tilde{T}$. This is consistent with the effective temperature for harmonic oscillator ($\chi=0$) given in Eq.~\ref{beta_eff}. For high temperatures, $\beta_{eff}={1}/{\tilde{T}}$. It is also interesting to note that for symmetric system-bath coupling, i.e, $\Gamma_1=\Gamma_2$, the effective temperature is just the mean temperature, $\tilde{T}=T_m=(T_1+T_2)/2$. However, the description in terms of an effective temperature is not possible for low temperatures, except when $\chi = 0$.

The high temperature scaling of the normalization constant can now be easily found by noting that for high temperatures, the summation can be converted into an integral. So we have
\begin{align}
\tilde{Z}\approx \int_0^\infty dx \exp\left[-\frac{1}{\tilde{T}}( \Omega_0 x +\chi x^2 )\right] \approx \frac{\sqrt{\pi}}{2}\sqrt{\frac{\tilde{T}}{\chi}}~.
\end{align}
The second step requires the condition $\tilde{T}\chi \gg \Omega_0^2$, under which  we see that  the normalization constant scales as $\sqrt{\tilde{T}/\chi}$.

\begin{figure}[t]
\includegraphics[scale=0.42]{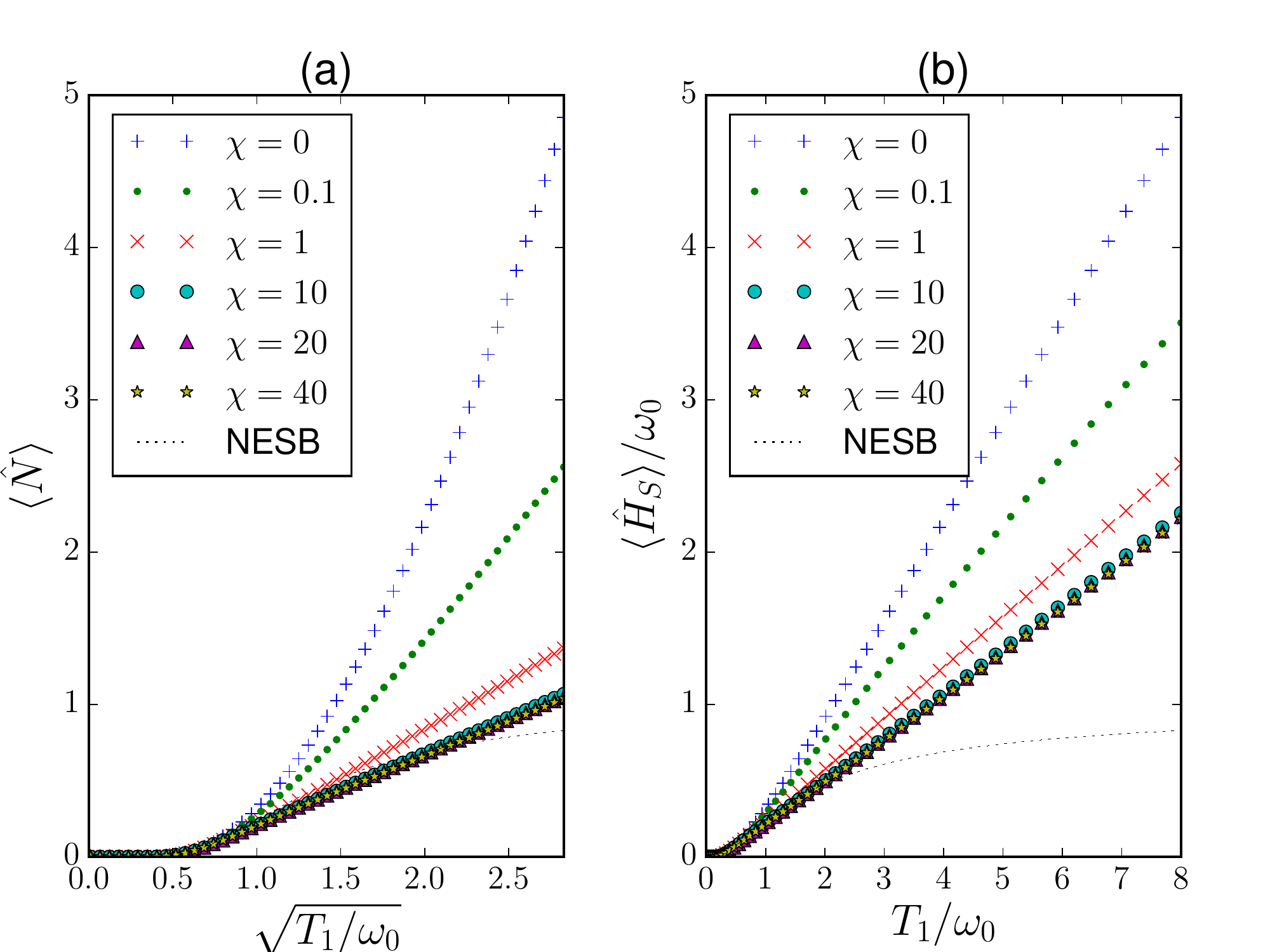} 
\caption{(color online) The plot shows scaling behaviour of average occupation and average energy of the single site Bose-Hubbard model for fixed $r={T_2}/{T_1}={1}/{3}$. Here $\omega_0 = \Omega_0+\chi$. The dotted plots correspond to the NESB model. For $\chi \gg \Omega_0$ ($\Omega_0=1$), there is data collapse over the entire range of temperature. Even for small $\chi (=0.1)$, there is substantial deviation from linear ($\chi=0$) behaviour. For high temperatures, $\langle \hat{N} \rangle$ scales as $\sqrt{T_1/\omega_0}$, while, $\langle \hat{H}_S \rangle$ scales as $T_1$. Here the system-bath coupling is taken symmetric : $\Gamma_1=\Gamma_2=1$. All energy variables are measured in units of $\Omega_0$.} \label{fig:occ_and_energy}
\end{figure} 

The above trick can be used to find high temperature scaling of average of any operator which is diagonal in the eigenbasis of the system Hamiltonian. The average of any operator $\hat{h}$ which is diagonal in the eigenbasis of the system has the form
\begin{align}
\label{avg_form}
\langle \hat{h}\rangle = \sum_{n=0}^\infty \rho_n h(n)=\frac{[h(0)+\sum_{n=1}^\infty e^{-\sum_{p=1}^{n} \log\left(f(\omega_{p-1}) \right)}h(n)]}{\tilde{Z}}~,
\end{align}
which, for $T_1,T_2\gg \Omega_0,\chi$,  exactly following above arguments, becomes
\begin{align}
\label{avg_form_highT}
\langle \hat{h}\rangle \approx \frac{\sqrt{2}}{\pi}\sqrt{\frac{\chi}{\tilde{T}}} \int_0^\infty dx \exp{\left[-\frac{ \Omega_0 x +\chi x^2 } {\tilde{T}}\right]} h(x)~.
\end{align}
Using this, we readily obtain the high temperature behaviour of  $\langle \hat{N} \rangle$ and $\langle \hat{H}_S \rangle$
\begin{align}
\label{occ_energy_highT}
\langle \hat{N} \rangle \approx \sqrt{\frac{\tilde{T}}{\pi \chi}}, \hspace{3pt} 
\langle \hat{H}_S \rangle \approx \tilde{T}+ (\Omega_0+2\chi) \sqrt{\frac{\tilde{T}}{\pi \chi}}~.
\end{align}
Thus $\langle \hat{N} \rangle$ should show a data collapse for various $\chi$ and vary as a function of $\sqrt{{\tilde{T}}/{\chi}}$, whereas, $\langle \hat{H}_S \rangle/\chi$ varies as ${\tilde{T}}/{\chi}$ and should show a data collapse for $\chi \gg \Omega_0$.

To check the above high temperature discussion and the connection with equilibrium behaviour, we now first define 
\begin{align}
r=\frac{T_2}{T_1}~,
\end{align}
which quantifies the degree of deviation from equilibrium, $r~=~1$ corresponding to  equilibrium. To ensure that the system is far from equilibrium, we keep $r$ fixed at $r<1$.  We note that, if $r$ is kept fixed, all the NESS results become a function of only one temperature, say, $T_1$. We choose $T_1/\omega_0$, where $\omega_0=\Omega_0+\chi$,  as the scaling variable since it can be used both in the highly interacting regime (where $\chi \approx \omega_0$) and the linear regime (where $\chi=0$). Also in the NESB regime, with $r$ fixed, all NESS quantities vary as a function of $T_1/\omega_0$, as can be checked from Eq.~\ref{rho_SB}.

In Fig.~\ref{fig:occ_and_energy}, we plot $\langle \hat{N} \rangle$ and $\langle \hat{H}_S \rangle/\omega_0$ as a function of $T_1/\omega_0$. Note that Eq.~\ref{occ_and_E}, and not any simplified expression, was used to calculate $\langle \hat{N} \rangle$ and $\langle \hat{H}_S \rangle$ in the plots. The plots clearly show data collapse over the entire temperature regime for $\chi \gg \Omega_0$. Also, important to note, is the substantial effect of small interaction strengths at high temperatures. The high temperature scaling behaviour for small interaction strengths is same as that for large interaction strengths, but there is no data collapse. This is because, the condition $\tilde{T}\chi \gg \Omega_0^2$ means that, for small $\chi$, there are large subleading terms. The low temperature behaviour matches with NESB. Since all these observations are valid for any choice of $r$, it follows that they are all valid in equilibrium (i.e, $r=1$) also. This is consistent with previous results in \cite{Haake1986}.

The crucial point in above discussion was to find that high temperature non-equilibrium results can be described via an effective temperature $\tilde{T}$. In the following, we will see that even transport properties at high temperatures can be described in terms of  $\tilde{T}$.

\subsection{\label{sec:F}Currents}

\begin{figure}[t]
\includegraphics[scale=0.45]{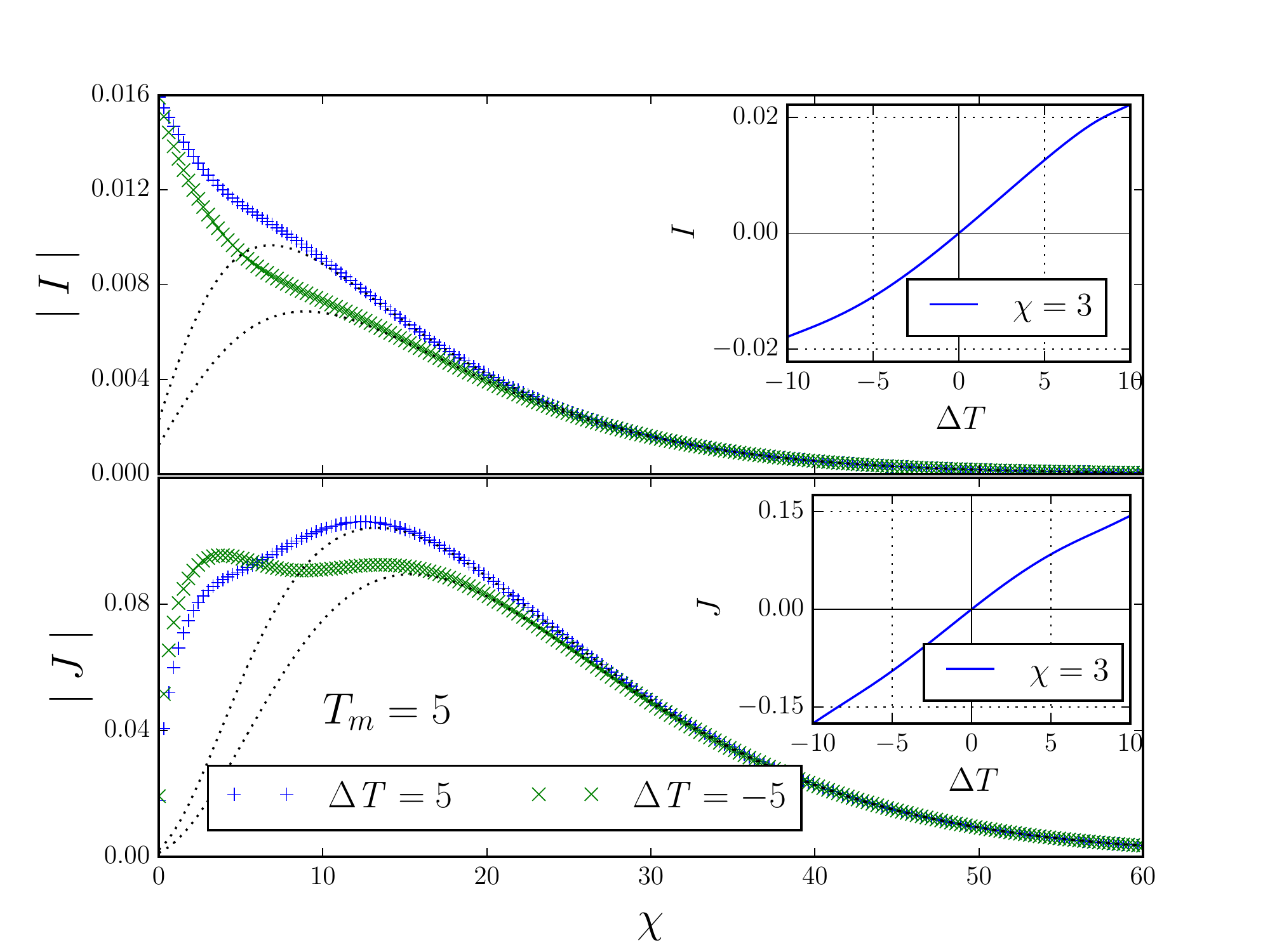} 
\caption{(color online) Plot of particle (top panel) and energy (bottom panel) currents as a function of interaction strength $\chi$ for transport under both forward and backward bias for Ohmic baths ($s=1$ in Eq.~\ref{J_general}) under asymmetric system bath coupling ($\Gamma_1 =0.4,\Gamma_2=1.6$). Here mean temperature $T_m=(T_1+T_2)/2=5$ for both plots.  The two black dotted lines in each plot correspond to the NESB currents for $\Delta T = \pm 5$. The currents from our model match the NESB currents for large $\chi$. Particle current decreases with $\chi$, while energy current behaves non-monotonically with $\chi$. The forward and backward currents do not match, thereby showing rectification effect. The direction of rectification of energy current is reversed beyond a value of $\chi$. It follows that at this value of $\chi$, the energy current does not show rectification. The insets show the corresponding currents as a function of $\Delta T= T_1-T_2$ for a chosen value of $\chi=3$ and $T_m=5$. It can be seen that $I$ and $J$ deviate from odd-function behaviour which is signature of rectification. 
Other parameters are $\varepsilon=0.1$, $\omega_c=1000$. All energy variables are measured in units of $\Omega_0$, and time is measured in units of $\Omega_0^{-1}$.  }
\label{fig:two_current}
\end{figure}

\begin{figure*}[t]
\includegraphics[scale=0.5]{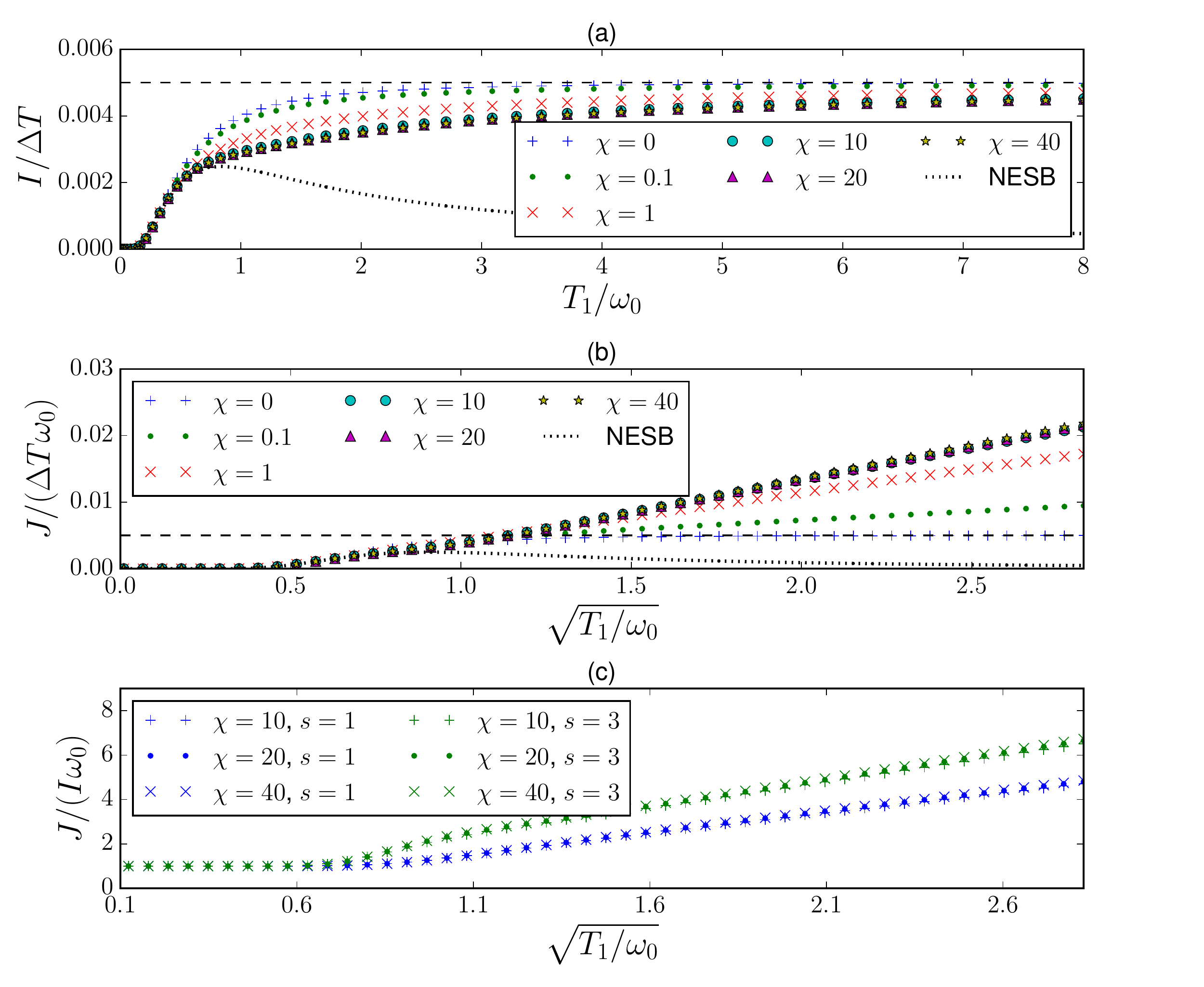} 
\caption{(color online) Panel (a) and (b) show the scaling behaviour of $I/\Delta T$ and $J/(\Delta T\omega_0)$ for Ohmic baths ($s=1$ in Eq.~\ref{J_general}) and for fixed $r=\frac{T_2}{T_1}=\frac{1}{3}$. The dotted lines show the corresponding NESB result. The horizontal dashed lines in panel (a) and panel (b) correspond to $I/\Delta T = J/(\Delta T\omega_0)=A={\varepsilon^2 \Gamma_1 \Gamma_2}/{(\Gamma_1+\Gamma_2)}$, which is the high temperature result for the harmonic oscillator ($\chi=0$).  For $\chi \gg \Omega_0$ ($\Omega_0=1$), there is a data collapse for all temperatures.  The NESB result matches with the non-equilibrium SSBH model for small temperatures. At higher temperatures, the NESB result shows non-monotonicity, which is not seen in the non-equilibrium SSBH model which demonstrates a stark difference between the two models. In panel (a), $I/\Delta T$ approaches a constant value at high temperatures irrespective of the strength of interaction strength. In panel (b), even for small interaction strength ($\chi=0.1$), substantial deviation from linear ($\chi=0$) behaviour is noticeable.  Panel (c) shows data collapse of $J/(I\omega_0)$ for all temperatures for $\chi \gg \Omega_0$. Irrespective of choice of bath spectral function and for fixed $r(=\frac{1}{3})$, $J/(I\omega_0)$ shows a data collapse and goes as $\sim \sqrt{T_1/\omega_0}$ for $T_1\gg \omega_0$. All observations are also valid in linear response regime (i.e, $r\approx 1$), and give the temperature scaling of conductance.  Other parameters are:  $\varepsilon=0.1$, $\omega_c=1000$, $\Gamma_1=\Gamma_2=1$. All energy variables are measured in units of $\Omega_0$, and time is measured in units of $\Omega_0^{-1}$.  } \label{fig:currents}
\end{figure*} 

Now we look at the average transport properties of the system, in particular we compute particle and energy currents.
To calculate current we look at the evolution equations of the expectation values $\langle \hat{N} \rangle$ and $\langle \hat{H}_S \rangle$ of $\hat{N}$ and $H_S$. Since $\hat{N}$ and $ \hat{H}_S$ are diagonal in the eigenbasis of the system Hamiltonoan we can directly obtain the evolution of their expectation values from Eq.~(\ref{rho_n_equation}). This gives
\begin{align}
\frac{d\langle \hat{N} \rangle}{dt} =& \sum_n n \frac{d\rho_n}{dt} =
 \varepsilon^2 \sum_n \Big (  \rho_n C_n - \rho_n D_n \Big)~, \label{N_eqn}\\
\frac{d\langle \hat{H}_S \rangle}{dt} = & \sum_n (\Omega_0 n +\chi n^2) \frac{d\rho_n}{dt} \nonumber \\ &= \varepsilon^2 \sum_n \Big ( \omega_n \rho_n C_n - \omega_{n-1} \rho_n D_n \Big)~. 
\end{align}
Collecting all terms depending on each bath separately, the above equations can be written like continuity equations of the forms $d\langle \hat{N}\rangle/dt=  I_{1} - I_{2}, 
d\langle \hat{H}_S \rangle/dt=  J_{1} - J_{2}$, 
where $I_{\ell}$ ($J_{\ell}$) is the particle (energy) current flowing into the system from $\ell$th bath. In steady state, $I_{1}=I_{2}=I$ and $J_{1}=J_{2}=J$. The steady state expressions for currents are :
\begin{align}
I& = \varepsilon^2\sum_{n=0}^{\infty} \rho_n (C^{(1)}_{n}-D^{(1)}_{n}) =\sum_{n=1}^{\infty} \rho_n n \mathcal{I}(\omega_{n-1})~,\nonumber \\
J& = \varepsilon^2\sum_{n=0}^{\infty} \rho_n (\omega_n C^{(1)}_{n}-\omega_{n-1} D^{(1)}_{n}) = \sum_{n=1}^{\infty} \rho_n \omega_{n-1} n \mathcal{I}(\omega_{n-1})~, \label{currents}
\end{align}
with 
\begin{align}
\label{curr_op}
\mathcal{I}(\omega_{n-1}) = \varepsilon^2\Big [ \frac{\Gamma_1\Gamma_2\mathcal{J}(\omega_{n-1})(\mathfrak{n}_1(\omega_{n-1})-\mathfrak{n}_2(\omega_{n-1})) }{\Gamma_1\mathfrak{n}_1(\omega_{n-1})+\Gamma_2\mathfrak{n}_2(\omega_{n-1})}   \Big ]~. 
\end{align}
The second steps of Eq.~\ref{currents} have been arrived at from the first steps after some simplification using the property of NESS density matrix given in Eq.~\ref{rho_n_condition}.
Note that energy and particle currents are not independent. But, in general, there is no way of directly finding one current given the other and they can have quite different behaviour. Fig.~\ref{fig:two_current} shows variation of energy and particle currents with interaction strength $\chi$ for Ohmic baths (($s=1$ in Eq.~\ref{J_general})), for both forward ($\Delta T >0$) and backward ($\Delta T <0$) biases. The mean temperature $T_m=(T_1+T_2)/2$ is kept fixed in the plots, and system-bath coupling is asymmetric ($\Gamma_1 \neq \Gamma_2$). The plots immediately show us a number of physical aspects of the system. 

Firstly, we note that the particle current decreases with increase in interaction strength $\chi$. This is expected because of increasing repulsive interaction in the system. On the other hand, energy current shows non-monotonic behaviour with $\chi$. This is plausible because, while, with increasing $\chi$, system allows less number of particles to pass, higher energy particles have a larger probability to pass through the system. 

Secondly, we see that there is rectification of both energy and particle currents, since the particle and energy currents for forward and backward biases do not match. This is to be expected because the expressions for currents in Eq.~\ref{currents} are not antisymmetric under interchange of hot and cold baths (i.e, $\mathfrak{n}_1\leftrightarrow \mathfrak{n}_2$) in general. It is only so under special conditions. Two of such special conditions where there is no rectification are when $\chi=0$, i.e, when the system is linear, and when $\Gamma_1=\Gamma_2$ (for any $\chi$). These can be easily checked from the expressions for currents (Eq.~\ref{currents}).  Hence, in general, there will be rectification effects in both particle and energy currents for $\chi\neq 0$ and $\Gamma_1 \neq \Gamma_2$.  
This is the generic behaviour in non-linear (interacting) systems.

Thirdly, as discussed before, for $\chi \gg \Omega_0, T_1, T_2$, the system behaves as NESB, and currents match with the NESB results. But, the rectification in the NESB limit is less than that for smaller interaction strengths. Thus rectification behaviour is non-monotonic as a function of $\chi$. Our findings therefore suggest that a careful engineering of the system Hamiltonian is required to get maximum rectification from a given system. 

Finally, and most interestingly, for small interaction strength, the rectification of energy current occurs in the opposite direction to rectification of particle current. Also, there is a non-zero value of $\chi$ where the forward and backward energy currents match, and hence there is no rectification. At this point, the system rectifies particle current but not the energy current. Beyond this value of $\chi$, energy and particle rectification occur in the same direction (see Fig. \ref{fig:two_current}, bottom panel).\\

In what follows, we investigate the behaviour of particle and energy currents and rectification in more detail along with their scaling behaviour.  

\subsubsection{Scaling behaviour of currents} 
As we have seen with average system properties, transport properties also behave differently for different relative values of temperatures and interaction strength. In the NESB regime, $\chi \gg \Omega_0, T_1, T_2$, the currents are given by
\begin{align}
\label{currents_SB}
&I_{SB} \approx   \frac{\varepsilon^2\Gamma_1\Gamma_2\mathcal{J}(\omega_{0})(\mathfrak{n}_1(\omega_{0})-\mathfrak{n}_2(\omega_{0})) } {\Gamma_1 (1+2\mathfrak{n}_1(\omega_0))+\Gamma_2 (1+2\mathfrak{n}_2(\omega_0))}~, \nonumber\\
&J_{SB} \approx \omega_0  I_{SB}~, 
\end{align}
with $\omega_0 = \Omega_0+\chi$. This is identical to the expression for current previously derived for NESB \cite{SegalPRL2005}. Note that in NESB regime, energy current is proportional to particle current. This is because, in this limit, transport is allowed through transfer of exactly one particle through the system, and that particle has energy $\Omega_0+\chi$. This is not valid beyond the NESB regime.

Now, let us look at the high temperature regime, $T_1, T_2 \gg \chi, \Omega_0$, where NESB results are not valid. We note that the expression for the currents in Eq.~\ref{currents} has the same form as Eq.~\ref{avg_form}. So, the high temperature trick in Eq.~\ref{avg_form_highT} can be readily applied to obtain    
\begin{align}
I^{\tilde{T}\gg \Omega_0,\chi} &\approx \sqrt{\frac{2\chi}{\pi^2\tilde{T}}} \int_0^\infty dx e^{-\frac{ \Omega_0 x +\chi x^2 } {\tilde{T}}} (x+1) \mathcal{I}(\omega_x) ~,\nonumber \\
&\approx\sqrt{\frac{2\chi}{\pi^2\tilde{T}}} \int_0^\infty dx e^{-\frac{ \Omega_0 x +\chi x^2 } {\tilde{T}}} (x+1) A \mathcal{J}(\omega_x) \frac{\Delta T}{\tilde{T}}~, \nonumber \\
J^{\tilde{T}\gg \Omega_0,\chi} &\approx\sqrt{\frac{2\chi}{\pi^2\tilde{T}}} \int_0^\infty dx e^{-\frac{ \Omega_0 x +\chi x^2 } {\tilde{T}}} (x+1)\omega_x A \mathcal{J}(\omega_x) \frac{\Delta T}{\tilde{T}}~,
\end{align}
with $A={\varepsilon^2 \Gamma_1 \Gamma_2}/{(\Gamma_1+\Gamma_2)}$ and $\Delta T = T_1 - T_2$. In the second step above, we have expanded the Bose distributions in $\mathcal{I}(\omega_n)$ (Eq.~\ref{curr_op}) for high temperatures to obtain $\mathcal{I}(\omega_n)\approx
A \mathcal{J}(\omega_n) {\Delta T}/{\tilde{T}}$. Using the general form of spectral function given in Eq.~\ref{J_general}, and after some algebra, we obtain
\begin{align}
\label{currents_highT}
I^{\tilde{T}\gg \Omega_0,\chi} &\approx K\big(s\big) = \frac{A\Delta T}{\sqrt{\pi}\tilde{T}}\int_0^{\infty} dy \Bigg[ y^{-\frac{1}{2}}e^{-y}\left(\sqrt{\frac{\tilde{T}y}{\chi}} + 1\right) \nonumber \\
&\big(\Omega_0+\chi+2\sqrt{\tilde{T}\chi y}\big)^s 
\Bigg]  ~,\nonumber \\
J^{\tilde{T}\gg \Omega_0,\chi} &\approx  K(s+1).  
\end{align} 
For the choice of spectral function in Eq.~\ref{J_general}, we can relate particle and energy currents via the function $K(s)$. We now look at the properties of the function $K(s)$. First, we look at the regime $\tilde{T}\gg \chi \gg \Omega_0$. In this regime, $K(s)$ becomes
\begin{align}
\label{K}
\frac{K\big(s\big)}{\chi^{s-1}\Delta T} &\approx \frac{A}{\sqrt{\pi}}\frac{\chi}{\tilde{T}}\mathcal{F}\big(\frac{\tilde{T}}{\chi},s\big)~, \\ 
{\rm where}~~\mathcal{F}(z,s)&= \int_0^{\infty} dy \Big[ y^{-\frac{1}{2}}e^{-y}\nonumber \\
&\times \big(\sqrt{zy} + 1\big) \big(1+2\sqrt{zy}\big)^s \Big]~. \nonumber 
\end{align}
This then gives
\begin{align}
\frac{K\big(s\big)}{\chi^{s-1}\Delta T} &\approx \frac{2^{s-1}A}{\sqrt{\pi}}\big(\frac{\tilde{T}}{\chi}\big)^{\frac{s-1}{2}}s\mathbf{\Gamma}\left(\frac{s}{2}\right) ,\hspace{10pt}\forall \hspace{5pt} {\tilde{T}\gg\chi\gg\Omega_0}~.
\end{align}
where $\mathbf{\Gamma}(x)$ is the Gamma function.
We immediately make the following observations. Firstly, for $T_1,T_2 \gg \chi \gg \Omega_0$, $K\big(s\big)/(\chi^{s-1}\Delta T)$ varies as a function of ${\tilde{T}}/{\chi}$. Thus, $I/\big(\chi^{s-1}\Delta T\big)$ and $J/\big(\chi^{s}\Delta T\big)$ for various values of $\chi$ should show a data collapse when plotted with ${\tilde{T}}/{\chi}$.  

Secondly, we see $K(1)\approx A\Delta T$. It follows that for Ohmic baths ($s=1$ in Eq.~\ref{J_general}) at high temperatures, particle current is independent of the interaction strength $\chi$ and one gets a linear-response-like relation $I\approx A\Delta T$, even for large temperature bias. This is consistent with high temperature result for a harmonic oscillator ($\chi=0$). However, the energy current $J\sim \sqrt{\tilde{T}/\chi} \Delta T $ and shows the effect of interaction. On the other hand for constant bath ($s=0$), the energy current $J$ always satisfies the linear-response-like relation, whereas the particle current is suppressed by a  factor $\sqrt{\chi/\tilde{T}}$.  

Thirdly, we note that the quantity $J/(\chi I)$ varies as a function of $\tilde{T}/\chi$ and scales as $\sqrt{\tilde{T}/\chi}$ for any $s$,
\begin{align}
\frac{J}{\chi I}\approx \frac{\mathcal{F}\big(\tilde{T}/\chi, s+1\big)}{\mathcal{F}\big(\tilde{T}/\chi, s\big)}&\approx {2}\sqrt{\frac{\tilde{T}}{\chi}} \hspace{2pt} \frac{(s+1)\mathbf{\Gamma}\left(\frac{s+1}{2}\right)}{s\mathbf{\Gamma}\left(\frac{s}{2}\right)}  \nonumber \\
&\hspace{10pt} \forall \hspace{5pt} T_1,T_2 \gg \chi \gg \Omega_0
\end{align}  
In the other regime, $T_1,T_2 \gg \Omega_0 \gg \chi$,  the scaling behaviour is the same, but the data collapse is difficult to see because of large sub-leading terms.

Note that, $I/\Delta T$ and $J/\Delta T$ actually give  the {\emph{beyond linear response}}  analog of particle and energy conductance. For linear response, $\tilde{T}=T$ in the RHS of Eq.~\ref{K}, $T$ being the equilibrium temperature. Then, the above discussion gives the high temperature scaling of conductance.

In Fig.~\ref{fig:currents}, we show the scaling behaviour of $I/\Delta T$ and $J/\Delta T$ as a function of $T_1/\omega_0$. As in Sec.~(\ref{sec:E}), we have chosen $T_1/\omega_0$ as the relevant scaling variable and have kept $r$ fixed at $r<1$ to ensure beyond linear response regime. $r\rightarrow 1$ gives the linear response conductance. The plots (a) and (b) of Fig.~\ref{fig:currents} are then the beyond-linear-response equivalent of temperature scaling of conductance. Note that Eq.~\ref{currents} (and not any simplified expression) was used to calculate currents. The plots show the scaling behaviour discussed above. The plots for $\chi \gg \Omega_0$ show data collapse over the entire range of temperatures. Also, the low temperature behaviour is given by NESB.
   
In Fig.~\ref{fig:currents}(a), we plot $I/\Delta T$ for fixed $r$ for Ohmic baths ($s=1$ in Eq.~\ref{J_general}).  We see that, for NESB, this quantity behaves non-monotonically with temperature, while for our non-equilibrium SSBH model, this quantity monotonically increases with temperature. In fact, the deviation from NESB result starts precisely at the point where the NESB result reaches a maximum. Similar behaviour is observed for $J/(\Delta T \omega_0)$ in Fig.~\ref{fig:currents}(b).  We conclude that, the non-monotonic behaviour of conductance of NESB, both at and beyond linear response, comes as a result of truncation of the energy spectrum and is not observed when all energy levels are considered. This demonstrates a major difference between the high temperature behaviour of non-equilibrium SSBH model and the NESB model. 

Also, it follows from Fig.~\ref{fig:currents}(a), that, for Ohmic baths ($s=1$ in Eq.~\ref{J_general}), at high temperatures the particle current (for fixed $\Delta T$) becomes independent of both interaction strength and the effective temperature. Fig.~\ref{fig:currents}(b) shows a considerable deviation from linear behaviour even for small interaction strengths. 

Fig.~\ref{fig:currents}(c) shows that, consistent with our previous discussion, irrespective of the choice of bath spectral function, the quantity $J/(I\omega_0)$, for $\chi \gg \Omega_0$, shows a data collapse and goes as $\sim \sqrt{T_1/\omega_0}$ for $T_1\gg \omega_0$ (with $r$ kept fixed).      

Having discussed the scaling behaviour of currents in detail, we now look into another important property of interacting systems, the rectification of current.

\begin{figure}[t]
\includegraphics[scale=0.42]{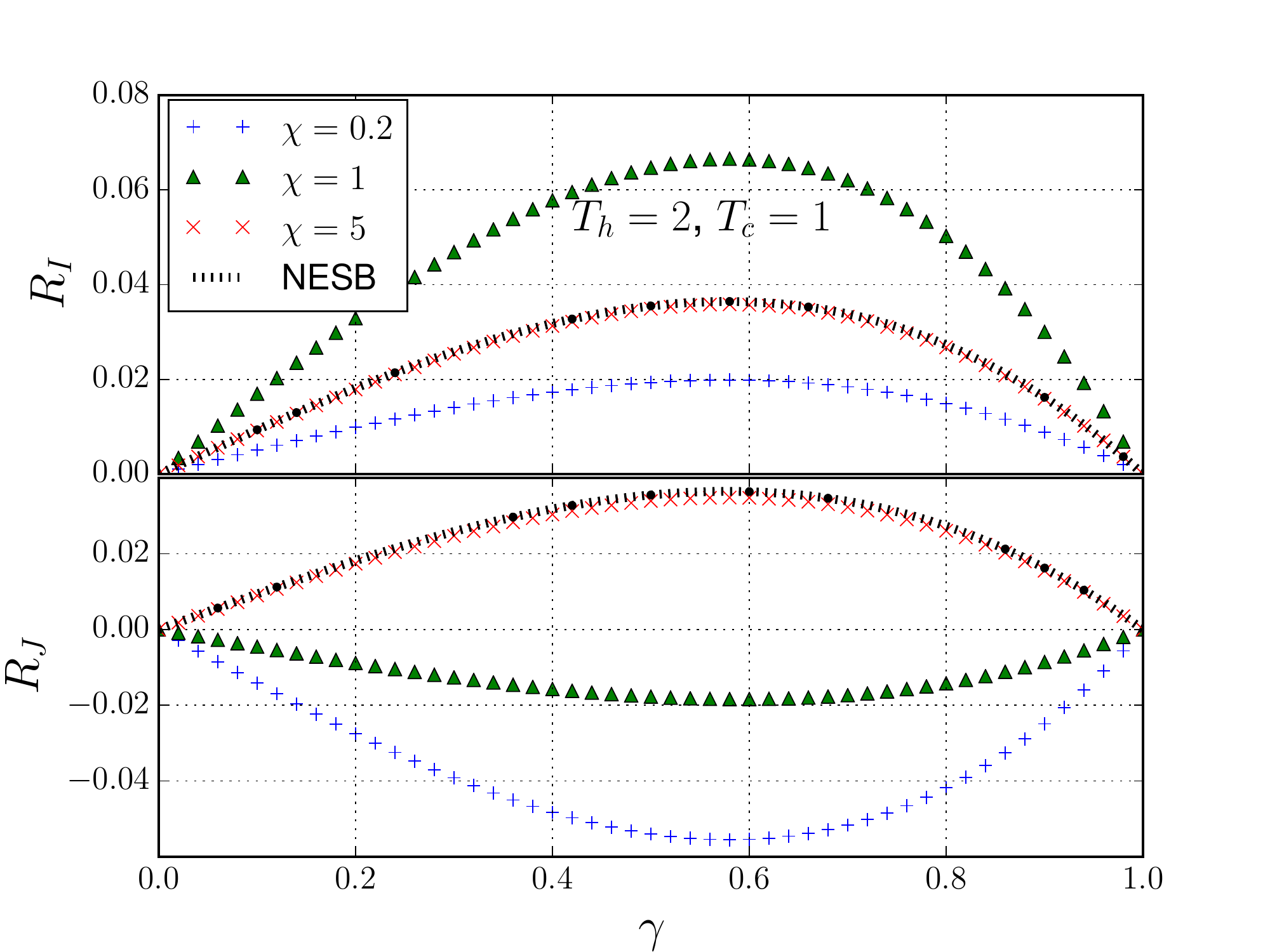} 
\caption{(color online) The plot shows particle rectification $R_I$ (top) and  energy rectification $R_J$ (bottom) as a function of asymmetry parameter $\gamma$ for various values of interaction strength $\chi$ and Ohmic baths ($s=1$ in Eq.~\ref{J_general}). $R_I$ and $R_J$ are as defined in Eq.~\ref{rectification_def}. The black dotted line corresponds to NESB. The rectfications become maximum when $\gamma\approx 0.6$.  Also, rectification of both particle and energy currents show a non-monotonic change with $\chi$. Rectification for large $\chi$ matches with the NESB result. Energy rectification changes direction with increase in $\chi$. Other parameters are as folows:  $\varepsilon=0.1$, $\omega_c=1000$. All energy variables are measured in units of $\Omega_0$, and time is measured in units of $\Omega_0^{-1}$.
}  \label{fig:three_rect_gamma}
\end{figure}

\subsubsection{Rectification}

Rectification of current is a generic behaviour of non-linear (interacting) systems in non-equilibrium.  As we have seen, in a non-equilibrium set-up, two kinds of currents through the system can be defined, the particle current and the energy current, and  their rectification behaviour can also be quite different. To our knowledge, there has been no previous work where both particle and energy current rectification for a bosonic non-linear system has been investigated. Also, note that rectification can only be observed beyond linear response regime.

Since rectification occurs only for asymmetric system-bath coupling, we use the following definition to describe the degree of asymmetry
\begin{align}
\label{Gamma}
\Gamma_1 = \Lambda(1-\gamma),\hspace{7pt} \Gamma_2 = \Lambda(1+\gamma)~,
\end{align}
where $0 \leq\gamma \leq 1$ is dimensionless. Given a value of asymmetry parameter $\gamma$, we define a measure of rectification as 
\begin{align}
\label{rectification_def}
R_I = \frac{I(\Delta T,\gamma)+I(-\Delta T,\gamma)}{I(\Delta T,\gamma=0)}, \hspace{3pt}
R_J = \frac{J(\Delta T,\gamma)+J(-\Delta T,\gamma)}{J(\Delta T,\gamma=0)}~.
\end{align}
$R_I$ and $R_J$ are the particle and energy current rectifications.  
This measure of rectification is as used in \cite{SegalPRL2005,SegalJChemPhys2005}. Note that, by this definition, rectification is positive if higher current flows when the cold bath is more strongly coupled to the system. Also, $R_I$ and $R_J$ are zero when $\gamma=0,1$. In our following discussion of rectification, we will primarily confine ourselves to the Ohmic baths ($s=1$ in Eq.~\ref{J_general}).

The variation of $R_I$ and $R_J$ with $\gamma$ is shown in Fig.~\ref{fig:three_rect_gamma}. Irrespective of the strength of interaction strength, we notice that the maximum rectification occurs when $\gamma\approx0.6$. The figure also shows, both particle and energy rectifications behave non-monotonically with $\chi$. For small $\chi$, $R_J$ is negative while $R_I$ is positive, hence, direction of energy rectification is opposite to particle rectification. For large $\chi$, the rectification is same as that obtained from NESB. In the NESB regime, particle and energy rectifications are same, because, particle current is proportional to energy current. All these observations are consistent with our discussion of Fig.~\ref{fig:two_current}.

\begin{figure}[t]
\includegraphics[scale=0.45]{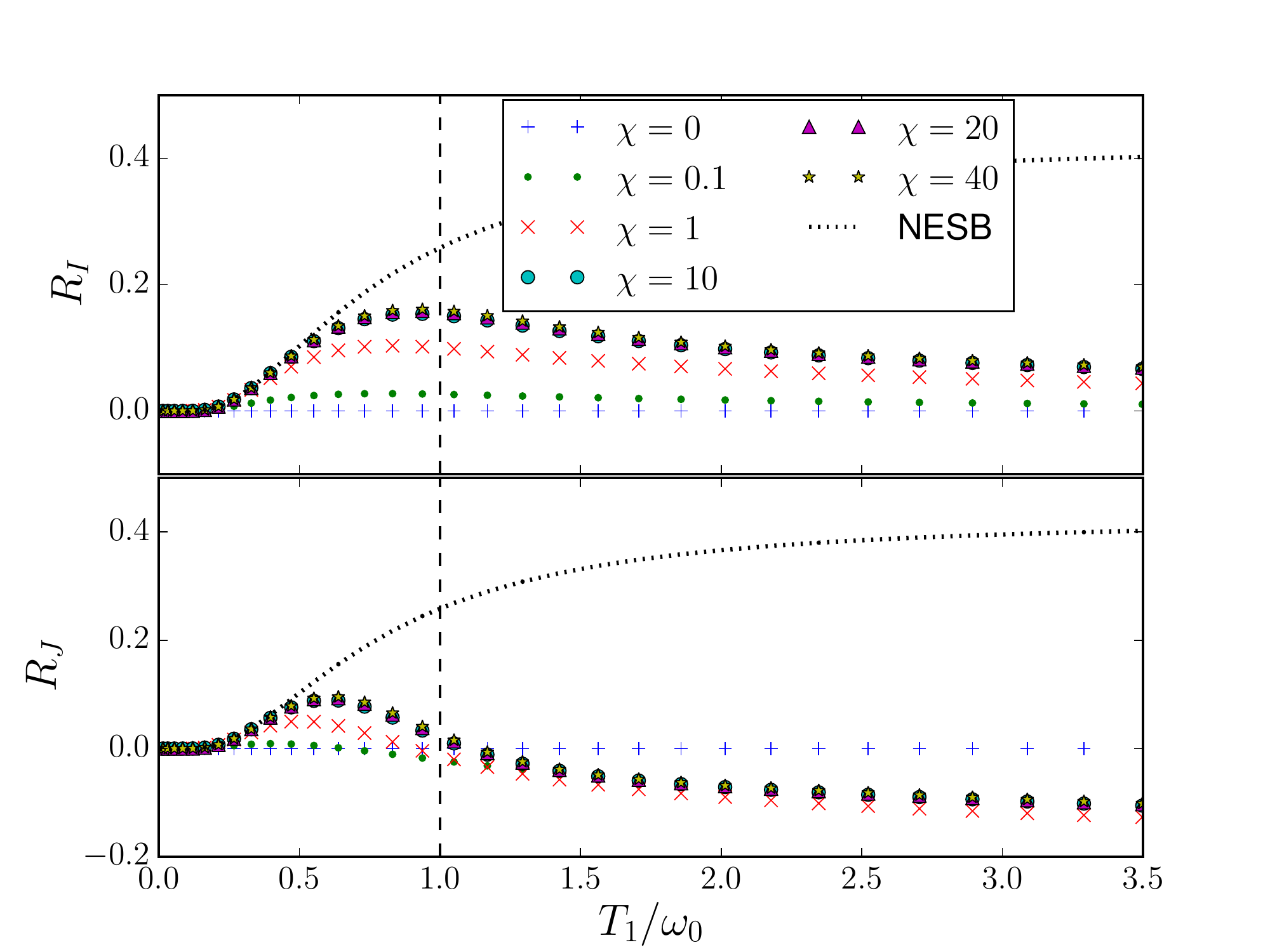} 
\caption{(color online) The plot shows rectification of particle current (top panel) and of energy current (bottom panel) of SSBH model out-of-equilibrium for fixed $r=\frac{T_2}{T_1}=\frac{1}{3}$, for Ohmic baths ($s=1$ in Eq.~\ref{J_general}). $\omega_0=\Omega_0+\chi$. $R_I$ and $R_J$ are as defined in Eq.~\ref{rectification_def}.  The veritcal dashed line indicates the positions of $T_1/\omega_0=1$. The dotted plots correspond to the NESB model. For $\chi \gg \Omega_0$, ($\Omega_0=1$), there is data collapse. Both $R_I$ and $R_J$ has a maximum for $\omega_0<T_1$. $R_J=0$ for $\omega_0\approx T_1$. $R_J$ shows reversal in direction of rectification beyond this point.  Other parameters are as folows:  $\varepsilon=0.1$, $\omega_c=1000$. All energy variables are measured in units of $\Omega_0$ and time is measured in units of $\Omega_0^{-1}$.}  \label{fig:rectification}
\end{figure}

To concisely investigate the rectification behaviour of the system as a function of the interaction strength and the temperatures, we again resort to the scaling variable $T_1/\omega_0$ with $r$ fixed at $r<1$. The plots of $R_I$ and $R_J$ as a function of $T_1/\omega_0$ for fixed $r$ for Ohmic baths  are shown in Fig.~\ref{fig:rectification}. We readily make the following observations :

Firstly, for $\chi < \Omega_0$, rectification is small. For $\chi \gg \Omega_0$, there is data collapse as expected from the scaling of currents.

Secondly, where NESB matches the non-equilibrium SSBH model, there is small rectification.  This can be understood from expression for currents in NESB regime given in Eq.~\ref{currents_SB}. NESB result holds when $\chi \gg \Omega_0,T_1,T_2$. Therefore, in this regime $\omega_0 \gg T_1,T_2$. So the bose distributions in Eq.~\ref{currents_SB} are exponentially small. Hence, $1+\mathfrak{n}(\omega_0) \approx 1$. With this approximation, the expressions of currents in Eq.~\ref{currents_SB} become antisymmetric under interchange of hot and cold baths. Thus the NESB regime of the SSBH model gives very small rectification. Maximum particle rectification $R_I$ is reached when $\omega_0 < T_1$ which is outside this regime. 

Thirdly, after the maximum, $R_I$ gradually approaches zero with increase in $T_1/\omega_0$. This is expected because, as we have seen before, at high temperatures, for Ohmic baths, the particle current eventually has the  form, $I\approx A \Delta T$, which is antisymmetric under interchange of hot and cold baths. On the other hand, the corresponding NESB rectification (the dotted line in Fig.~\ref{fig:rectification}) continues to increase with $T_1/\omega_0$, until it saturates to a high value. Thus, as an effect of having all energy levels, and not truncating at two levels, the particle rectification is suppressed at high temperatures.  

Fourthly, and most interestingly, energy rectification $R_J$ also has a peak  for $\omega_0 < T_1$. However, $R_J=0$ at $\omega_0 \approx T_1$. At this point, the particle rectification is still positive. So the system behaves as a particle rectifier and not as an energy or heat rectifier. Beyond that point, $R_J$ changes sign. Thus the direction of rectification is reversed. With further increase in $T_1/\omega_0$,  energy rectification continues to grow in the reversed direction. Therefore, at high temperatures, the heat rectification occurs in the opposite direction to particle rectification, and continues to grow as temperatures of both hot and cold baths are increased.

Note that all the above observations are for Ohmic baths ($s=1$ in Eq.~\ref{J_general}). Finally, we discuss the case of non-Ohmic baths. For superohmic baths (i.e, for $s>1$), both particle and energy currents show a reversal of direction of rectification. The reversal of energy rectification for superohmic bath occurs at a lower value of $T_1/\omega_0$ than that for the Ohmic bath. For subohmic baths (i.e, for $s<1$), the reversal of energy rectification occurs at higher values of $T_1/\omega_0$ than that for the Ohmic bath. The particle rectification does not show reversal for subohmic baths. For the constant bath, $s=0$, neither particle nor energy rectification shows reversal.  However, in all cases, both particle and energy rectifications vary non-monotonically with the interaction strength.

\subsection{\label{sec:TD}Time dynamics}

Until now, we have discussed the properties of the NESS of the out-of-equilibrium SSBH model. In this section, we look at the transient time dynamics of the various physical quantities we have so far calculated in NESS.

To do this we revert to the equation for time evolution of $\rho_n$, Eq.~\ref{rho_n_equation}. The equation can be re-written and solved in the form
\begin{align}
\label{rho_time}
\frac{\partial\tilde{\rho}(t)}{\partial t}= -\varepsilon^2 M \tilde{\rho}(0) \Rightarrow \tilde{\rho}(t) = e^{-\varepsilon^2 M t}\tilde{\rho}(0)
\end{align}
where $\tilde{\rho}$ is  an infinite dimensional column vector containing diagonal elements of the density matrix and  
$M$  is a infinite dimensional non-Hermitian square matrix containing the entries of Eq.~\ref{rho_n_equation}. $M$ has the form
\begin{align}
M=
\begin{bmatrix}
C_0+D_0 & -D_1 &0 &\ldots &\ldots &\ldots\\
-C_0 & C_1+D_1 & -D_2 &0 &\ldots &\ldots\\
0 &-C1 & C_2+D_2 &-D_3 &0 &\ldots\\
\vdots &\ddots &\ddots &\ddots &\ddots &\ddots
\end{bmatrix}
\end{align} 
where $C_n$ and $D_n$ are as defined in Eq.~\ref{CD}. Note that the matrix $M$ has the form of a Markov matrix. The sum of each column is zero ($D_0=0$ by definition). This corresponds to the fact that the trace of the density matrix is preserved, i.e, $\sum_n \rho_n = 1$.

To calculate the time dynamics, we choose  an initial state with no particles in the system, i.e, initially, $\rho_0(0)~=~1$, and $\rho_n(0)=0$ $\forall n \neq 0$. The Eq.~\ref{rho_time} is used to numerically obtain the time evolution. Even though the matrices involved are infinite dimensional, for given interaction $\chi$ and temperatures $T_1$ and $T_2$, only a finite number of levels, determined by the ratio of the temperatures and the interaction, effectively contribute. Thus, starting from a finite matrix size, a convergence is reached as the matrix size is increased. Smaller interaction and higher temperatures require larger matrix sizes. A subtle point to note is that, if the matrix $M$ is truncated at any finite size, say $p$, then the constraint that the sum of each column should be zero is not satisfied for the $p$th column, unless $C_p=0$. Consequently, the matrix $M$ can be truncated at size $p$ only if $C_p \ll D_p$.

Since we are using Redfield equation under Born-Markov approximation to obtain transient time dynamics, we need to be careful in choosing the observation times. This is because, as mentioned in Sec.~\ref{sec:C}, Markov approximation is valid only when observation times are much larger than the time for decay of bath correlation functions. We have made some  estimates (similar to that in Ref. \cite{archak}), which indicate that the bath relaxation times are indeed much smaller than the  observational transient times of our interest.

\begin{figure}[t]
\includegraphics[scale=0.42]{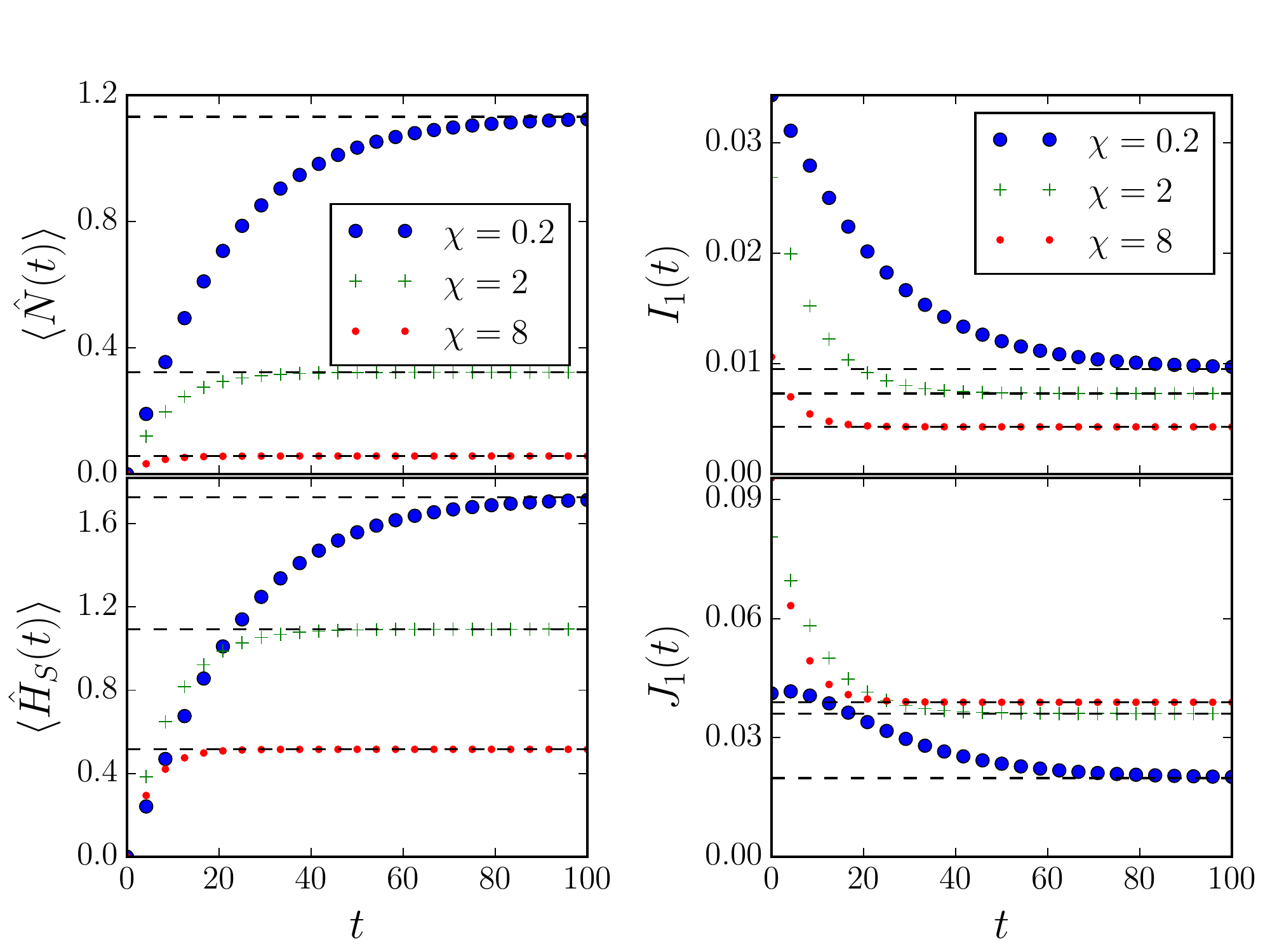} 
\caption{(color online) The figure shows time dynamics of (a) average occupation, $\langle N(t) \rangle$,(b) average energy, $\langle H_s (t) \rangle$, (c) particle current from left bath, $I_1(t)$, and (d) energy current from left bath, $J(t)$, for various values of interaction strength and for a temperature bias $T_1=4$ and $T_2 =2$ for Ohmic baths ($s=1$ in Eq.~\ref{J_general}). All physical quantities demonstrate a non-unitary evolution towards a steady state. The approach to steady state is faster for higher interactions. Since physical quatities plotted here are diagonal in the eigenbasis of the system Hamiltonian, none of them shows oscillations with time. Other parameters are $ \Omega_0~=~1, \varepsilon~=~0.1, \Gamma_1~=~0.4, \Gamma_2~=~1.6, \omega_c~=~1000$. All energy variables are measured in units of $\Omega_0$, and time is measured in units of $\Omega_0^{-1}$.  }  \label{fig:dyn}
\end{figure}

Time dynamics of physical quantities $\langle \hat{N}(t) \rangle $, $\langle \hat{H}_s (t)\rangle $, $I(t)$, $J(t)$ for Ohmic baths ($s=1$ in Eq.~\ref{J_general}) are shown in Fig.~\ref{fig:dyn}, where $I_1(t)$ and $J_1(t)$ are respectively particle and energy currents from the left bath into the system. (Before reaching steady state, the currents from left and right baths are not the same.) Similar time dynamics, but for bosonic system of two sites without interactions, was calculated in Ref.~\cite{archak}.   We observe that, unlike the two site non-interacting case in Ref.~\cite{archak}, here, the currents show no oscillations. More interestingly, we also observe that the time to reach steady state, called $t_{ss}$ hereafter, decreases with increase in the interaction. In the following, we investigate the dependence of $t_{ss}$ on set-up parameters more carefully.

\begin{figure}[t]
\includegraphics[scale=0.45]{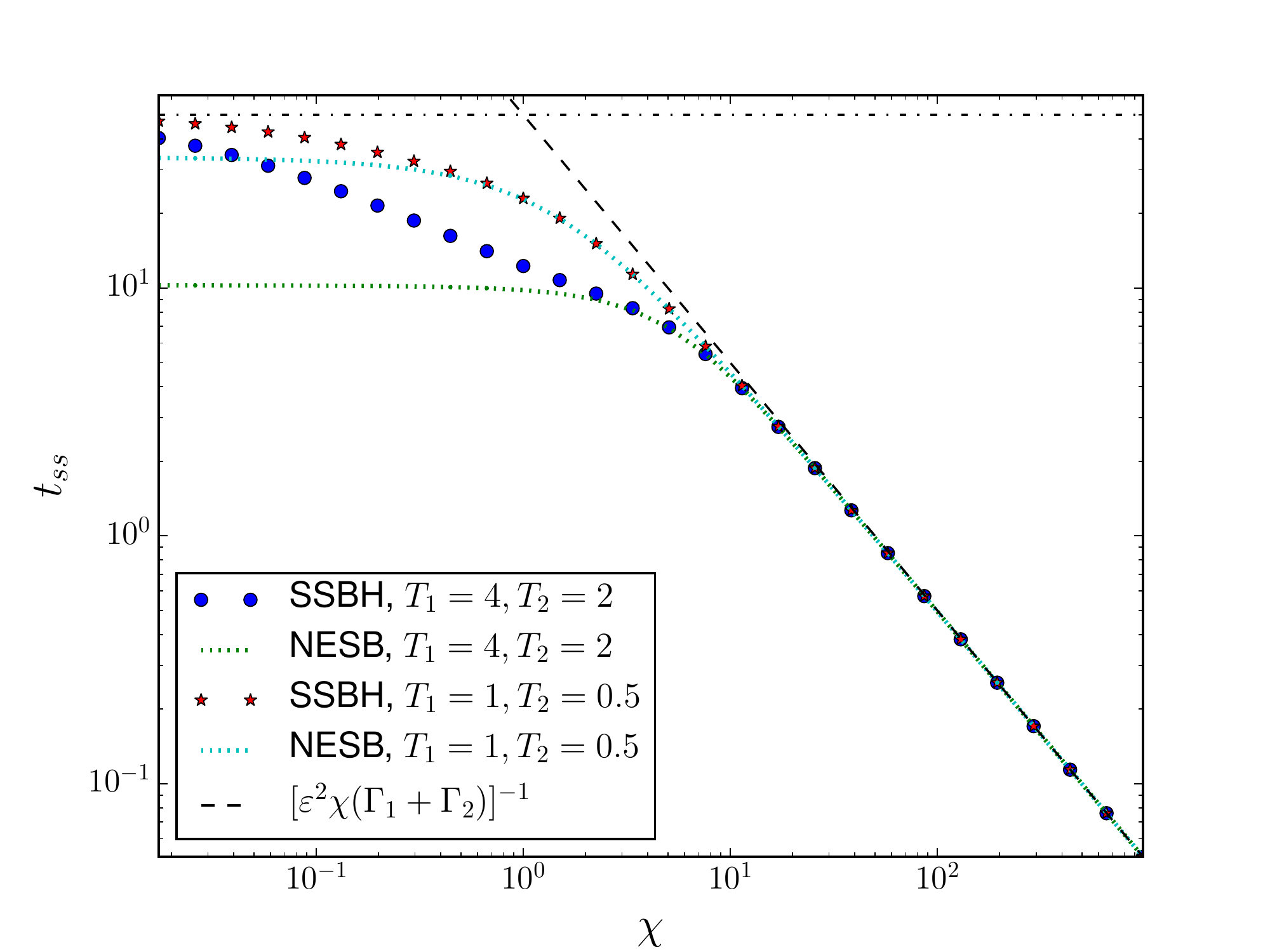} 
\caption{(color online) The figure shows a log-log plot of time to reach NESS, $t_{ss}$ (as defined in Eq.~\ref{t_ss_def}) as a function of interaction strength $\chi$ for Ohmic baths ($s=1$ in Eq.~\ref{J_general}). The horizontal dash-dotted line corresponds to $\Big[\varepsilon^2\sum_{\ell=1}^2\mathcal{J}_\ell(\Omega_0)\Big]^{-1}$ which is the $t_{ss}$ for the linear ($\chi=0$) system. For large $\chi$, $t_{ss} \approx \Big[\varepsilon^2 \chi \left( \Gamma_1+\Gamma_2\right) \Big]^{-1}$, and this is indicated by the dashed line. For intermediate $\chi$, $t_{ss}$ depends on the temperatures of the hot and cold baths. Other parameters are $ \Omega_0=1, \varepsilon=0.1, \Gamma_1=0.4, \Gamma_2=1.6, \omega_c=1000$. All energy variables are measured in units of $\Omega_0$, and time is measured in units of $\Omega_0^{-1}$. 
}  \label{fig:dyn2}
\end{figure} 
We note from the solution of $\tilde{\rho}(t)$ in Eq.~\ref{rho_time} that time can be scaled as $\varepsilon^2 t$. It follows that $t_{ss} \propto \varepsilon^{-2}$. Thus, time to reach steady state increases as system-bath coupling becomes weaker. 

It is also clear from Eq.~\ref{rho_time} that the steady state is given by the eigenvector of the matrix $M$ corresponding to zero eigenvalue. The fact that a unique steady state is reached in long time then implies that all other eigenvalues of the $M$ have positive real part. Interestingly, we have found in our numerical computation that the eigenvalues of $M$ are all real (and hence no oscillations in time). The smallest non-zero eigenvalue then gives a measure of $t_{ss}$. So, we define
\begin{align}
\label{t_ss_def}
t_{ss}\equiv \frac{1}{\varepsilon^2\lambda_1}~,
\end{align}     
$\lambda_1$ being the smallest non-zero eigenvalue of $M$.

Even though $t_{ss}$ cannot be calculated for all set-up parameters analytically, in two limiting cases, analytical results can be obtained. The first case corresponds to the NESB regime $\chi \gg \Omega_0,T_1, T_2$. In this case, $C_1 \ll D_1$, only two levels effectively contribute and matrix $M$ has the form
\begin{align}
M\approx
\begin{bmatrix}
C_0 & -D_1 \\
-C_0 & D_1 
\end{bmatrix}
\end{align} 
The non-zero eigenvalue of $M$ is
\begin{align}
\lambda_1 = [\varepsilon^2 t_{ss}]^{-1} \approx C_0+D_1 = \sum_{\ell=1}^2 \mathcal{J}_\ell(\omega_0)\left( 2\mathfrak{n}_\ell(\omega_0)+1 \right)
\end{align} 
The time to reach steady state therefore also depends on the temperatures of the baths.
However, for large $\omega_0=\Omega_0+\chi$, the Bose distributions are expotentially small.
Thus, for $\chi \gg \Omega_0,T_1,T_2$ and for general bath of the form given in Eq.~\ref{J_general}, 
\begin{align}
t_{ss} \approx \frac{1}{\varepsilon^2 \chi^{s} \left( \Gamma_1+\Gamma_2\right)}~, 
\end{align}
which is independent of the temperatures of the baths. We also see that, for constant baths ($s=0$ in Eq.~\ref{J_general}), $t_{ss}$ is independent of interaction strength. For other baths, $t_{ss}$ decreases with increase of interaction strength as a power law.

The second case where $t_{ss}$ can be analytically, calculated corresponds to the linear system, $\chi=0$. In this case, $C_n$ does not decay with $n$ (Eq.~\ref{CD} for $\chi=0$) and hence the matrix $M$ cannot be truncated at any finite size. So, the above method of finding $t_{ss}$ fails. However, since, in this case, we have a non-interacting system, we can find $t_{ss}$ directly from the evolution equation for $\langle \hat{N}(t) \rangle$. The evolution equation for $\langle \hat{N}(t) \rangle$ can be obtained from Eq.~\ref{N_eqn} by setting $\chi=0$. The resulting equation can be written and solved in the form
\begin{align}
&\frac{d\langle \hat{N}(t) \rangle}{dt} = -\varepsilon^2 \langle \hat{N}(t) \rangle \sum_{\ell=1}^2 \mathcal{J}_\ell(\Omega_0) +\varepsilon^2 \sum_{\ell=1}^2 \mathcal{J}_\ell(\Omega_0)\mathfrak{n}_\ell(\Omega_0) \nonumber \\
&\Rightarrow \langle \hat{N}(t) \rangle = \left( \langle \hat{N}(0) \rangle-N_{ss}\right) e^{-\varepsilon^2 t\sum_{\ell=1}^2\mathcal{J}_\ell(\Omega_0)} + N_{ss} 
\end{align} 
with $N_{ss}= \Big[\sum_{\ell=1}^2 \mathcal{J}_\ell(\Omega_0)\mathfrak{n}_\ell(\Omega_0)\Big]/\Big[\sum_{\ell=1}^2 \mathcal{J}_\ell(\Omega_0)\Big]$ being the NESS occupation. From above equation, it is clear that for $\chi=0$, the $t_{ss}$ is given by
\begin{align}
t_{ss}=\frac{1}{\varepsilon^2\sum_{\ell=1}^2\mathcal{J}_\ell(\Omega_0)}
\end{align}
This is again independent of the temperatures of the baths.

Except for these two limiting cases, $t_{ss}$ needs to be found numerically using the definition Eq.~\ref{t_ss_def}. Fig.~\ref{fig:dyn2} shows log-log plot of numerically obtained $t_{ss}$ as a function of interaction strength $\chi$ for Ohmic baths for two different choices of temperatures of hot and cold baths. It is seen that except for the limiting cases, $t_{ss}$ depends on the temperatures of the baths. The limiting cases show the behaviour discussed above.

\subsection{\label{sec:G}Conclusion}

We have investigated a system consisting of a single bosonic site with Bose-Hubbard interaction weakly coupled to two bosonic baths at different temperatures. We have used the Redfield QME method to obtain analytical results beyond linear response regime. Below, we summarize our main findings and their potential applications.

We have found an analytical result for the population density (diagonal elements of the density matrix in the eigenbasis of system Hamiltonian) of the system in NESS (Eq.~\ref{rho_n}). This has further allowed us to find various physical observables like average occupation and energy, as well as the particle and energy currents in NESS. We have then analytically found interesting scaling behaviour of the physical observables. Our main finding in this respect is that the high temperature behaviour of the system can be described in terms of an effective temperature (Eq.~\ref{Teff}). Then, it follows that, with the ratio $r$ of the temperatures of cold and hot baths fixed, there occurs a data collapse for various strengths of interaction $\chi$, when physical observables of the system are plotted in terms of the scaling variable $T_1/\omega_0$ (Figs.~\ref{fig:occ_and_energy},\ref{fig:currents},\ref{fig:rectification}). The scaling behaviours hold for a general choice of bath spectral functions of the form given in Eq.~\ref{J_general}. We have also found very interesting rectification behaviour of the system.  The most interesting finding is that, for Ohmic and subohmic baths ($0<s\leq 1$ in Eq.~\ref{J_general}), the energy current shows a reversal in direction of rectification (Fig.~\ref{fig:rectification}). It follows that there is a non-zero strength of interaction strength, ($\chi\approx T_1 - \Omega_0$, for Ohmic baths), where energy or heat rectification is zero. At this point, the system behaves as a particle rectifier but not as a heat rectifier. For superohmic baths ($s > 1$ in Eq.~\ref{J_general}), both particle and energy currents show reversal in direction of rectification. For constant baths ($s=0$ in Eq.~\ref{J_general}), there is no change in direction of rectification for both particle and energy currents. Therefore careful engineering of baths can lead to various interesting rectification behaviour of the system. Such phenomena can be potentially used to create quantum devices, such as, optical diodes.  Reversal of direction of thermal rectification of a quantum system has also been previously theoretically seen for a Heisenberg spin chain out of equilibrium \cite{reversal}. Further, we have computed non-unitary time-dynamics of various physical quantities (Fig.~\ref{fig:dyn}). We found that, except for constant baths, the time to reach steady state is shorter for higher interactions and higher system-bath couplings. For constant baths, with increase in interaction, the time to reach steady state approaches a constant independent of the strength of interaction. For large $\chi$, the time to reach steady state goes as $t_{ss}\sim \big[ \varepsilon^2\chi^{s} \big]^{-1}$ for a general choice of bath spectral functions of the form given in Eq.~\ref{J_general}. All our results also are consistent with the linear, $\chi=0$ (harmonic oscillator) case, as well as the NESB, $\chi \gg \Omega_0, T_1,T_2$, case. 

Our results are experimentally relevant in quantum hybrid systems, where a single site Bose-Hubbard model can be realized,  as well as, in molecular junction systems, where our set-up describes a model for anharmonic junctions.  Future work includes going to strong system-bath coupling regime, as well as to generalize to non-equilibrium interacting systems of more than one site, such as, the non-equlibrium Bose Hubbard chain \cite{kordas,biella,hartmann1,hartmann2}  and Jaynes-Cummings Hubbard model \cite{JCH}. 

\textit{Acknowledgments:}
AD would like to thank  support   from the Indo-Israel joint research project No. 6-8/2014(IC) and from the  French Ministry of Education through the grant ANR (EDNHS)​.

\bibliographystyle{unsrt}

\end{document}